\documentstyle[11pt,moriond,epsfig]{article}

\input{psfig}

\def\Journal#1#2#3#4{{#1} {\bf #2}, #3 (#4)}

\def\be{\begin{equation}}
\def\ee{\end{equation}}
\def\bea{\begin{eqnarray}}
\def\eea{\end{eqnarray}}

\def\ltsima{\; \buildrel < \over \sim \;}
\def\simlt{\lower.5ex\hbox{\ltsima}}            
\def\gtsima{\; \buildrel > \over \sim \;}
\def\simgt{\lower.5ex\hbox{\gtsima}}            

\def\grs{\mbox{GRS } 1915+105}

\def\Kkmsdeg2{\mbox{ K km s}^{-1} \mbox{deg}^{2}}
\def\metre{\mbox{ m}}
\def\micrometre{\mbox{ } \mu \mbox{m}}

\def\arcsec{''}

\def\GeV{\mbox{ GeV}}
\def\ergs{\mbox{ erg.s}^{-1}}
\def\mag{\mbox{ magnitude}}

\begin{document}
\vspace*{4cm}
\title{INFRARED JET OF THE GALACTIC SUPERLUMINAL SOURCE \\
GRS 1915+105}

\author{ S. CHATY \& I.F. MIRABEL }

\address{CE Saclay, Service d'Astrophysique, DSM/DAPNIA/SAp, \\
L'Orme des Merisiers, Bât. 709, \\
F-91 191 Gif-sur-Yvette, Cedex, FRANCE}

\maketitle\abstracts{
A near-infrared jet emanating from the galactic superluminal source $\grs$, discovered at $\lambda = 2.2 \, \mu$m by Sams {\it et al.}~\cite{sam96} on 18-21 July 1995, was not seen later anymore, neither in the Mirabel {\it et al.}~\cite{mir96b} images, taken on 4 August 1995 and that we analyze here, nor by Eikenberry \& Fazio~\cite{eik96} on 16-17 October 1995. Therefore, the radiative lifetime of the electrons which have created this emission is $\tau < 17 \mbox{ days}$. The detection, and the later non-detection, of this infrared jet is consistent with a synchrotron emission, provided that the magnetic field in the ejected clouds is $B \gtsima 160$ mG, and that the Lorentz factor of the internal motions of the electrons in the jet is $\gamma \ltsima 10^{4}$. The lower limit for the magnetic field suggests a magneto-hydrodynamic mechanism for the acceleration and collimation of the ejecta.}

\section{The Observations}

Sams {\it et al.}~\cite{sam96} observed the counterpart of the galactic superluminal source GRS 1915+105, by near-infrared speckle imaging, at $\lambda = 2.2 \micrometre$ (K band). These observations were done on 18--21 July 1995, using the Max Planck Institut für Extraterrestrische Physik near-infrared speckle camera SHARP at the New Technology Telescope of the European Southern Observatory. The main characteristics, of all the instruments and observations reported in this paper, are given in the Table \ref{characteristics}. The high-spatial resolution images show that $\grs$ appears extended to the South-West: this jet has a total near-infrared magnitude of $K = 13.9 \mag$, and is separated from the central source by $0.3 \arcsec$. No extension is seen to the North-East. The jet structure, the flux ratio between the SW jet and the NE residuals, and the position angle of the SW jet, are consistent to that observed in a radio outburst of the source (Mirabel \& Rodríguez~\cite{mir94}).

Three months later, on 16 and 17 October 1995, Eikenberry \& Fazio~\cite{eik96} observed the same source $\grs$ in the K-band, using the COB infrared imager on the Kitt Peak National Observatory 2.1 m telescope. They found no evidence for extended structure in their combined PSF-subtracted images, placing an upper limit on any similar point-like emission of $K > 16.4\mag$, with a $95\%$ confidence level. They also searched for possible extended emission from the jet, and place a limit of $K > 17.7\mag$, with a $95\%$ confidence level.

To understand what happened between these two dates, it was necessary to have images that were taken in this period. Mirabel {\it et al.}~\cite{mir96b}, discovering in the infrared wavelengths the reverberation of an energetic outburst of $\grs$, observed this source on 4 August 1995, with the IRAC2b camera mounted at the F/35 infrared adapter of the ESO/MPI 2.2 m telescope at the European Southern Observatory. We have reanalyzed these images, with the treatments described in the following paragraphs.

\begin{table}[t]
\caption{The characteristics of the instruments and of the observations of $\grs$. \label{characteristics}}
\vspace{0.4cm}
\begin{center}
\begin{tabular}{|l|l|l|l|}
\hline
 & {\bf Sams {\it et al.}~\cite{sam96}} & {\bf Mirabel {\it et al.}~\cite{mir96b}} & {\bf Eikenberry \& Fazio~\cite{eik96}} \\
\hline
%
Date & 18--21 July 1995 & 04 August 1995 5.0 UT & 16--17 October 1995 \\
Julian Date & 2\,449\,916.5 & 2\,449\,933.7 & 2\,450\,006.5 \\
Telescope & ESO/NTT 3.5 m & ESO/MPI 2.2 m & KPNO 2.1 m \\
Instrument & SHARP & IRAC2b & COB IR\\
Field of View & $12.8'' \times 12.8''$ & $2' \times 2'$ & $40'' \times 40''$\\
Resolution & $0.05''/ \mbox{ pixel}$ & $0.49''/ \mbox{ pixel}$ & $0.2''/ \mbox{ pixel}$ \\
Seeing & $\gtsima 0.6''$ & 0.7'' & 0.7--1.1'' \\
Total K-magnitude & $12.5 \mag$ & $13.4 \mag$ & $13.5 \mag$ \\
Absolute flux error & $0.1 \mag$ & $0.1 \mag$ & $0.1 \mag$ \\
%
\hline
\end{tabular}
\end{center}
\end{table}

\begin{figure}

\setlength{\unitlength}{1.0cm}

\begin{picture}(6,4)(0.9,7)
\includegraphics{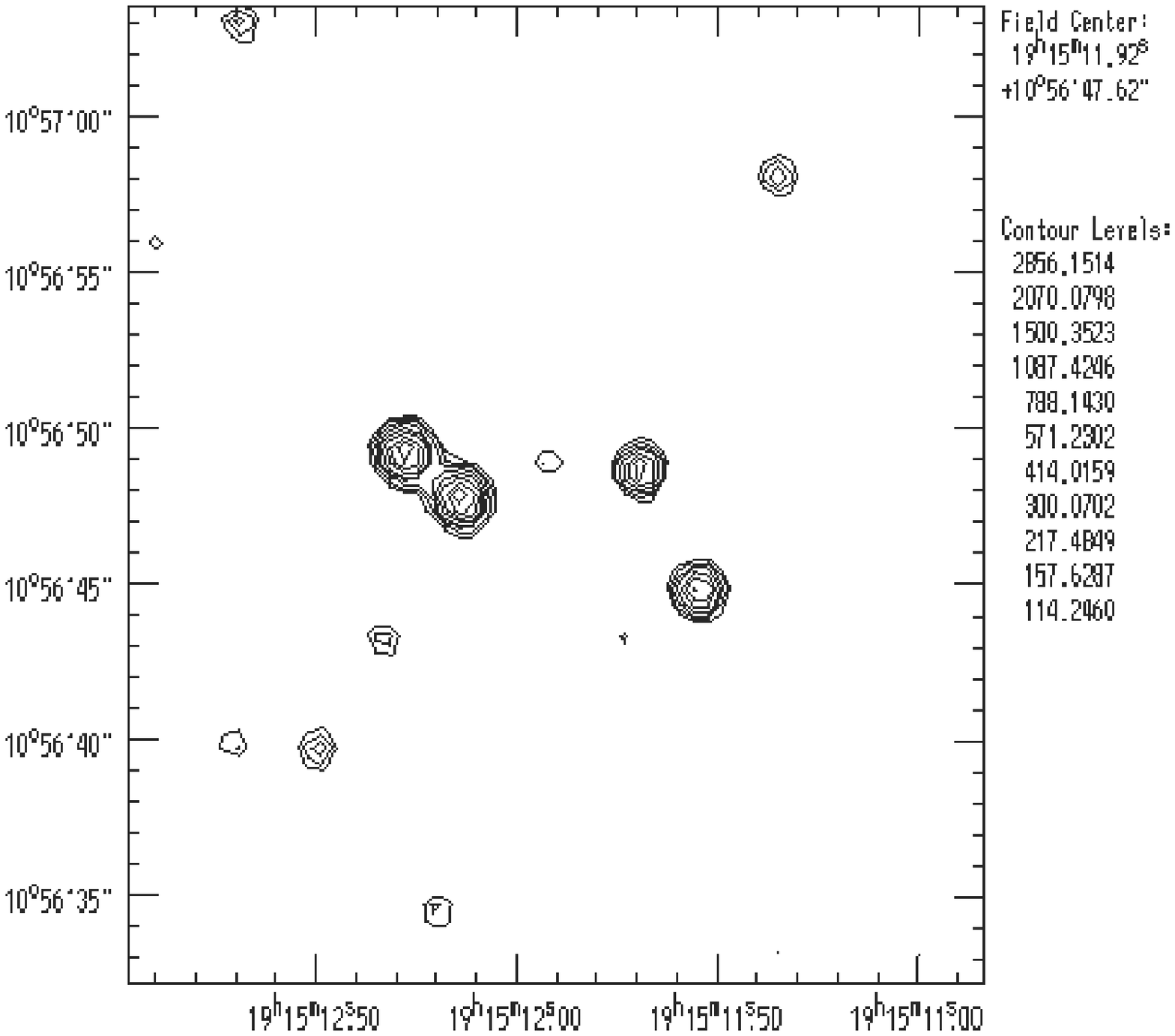}
\end{picture}
 
\begin{picture}(6,4)(-9,3)
\includegraphics{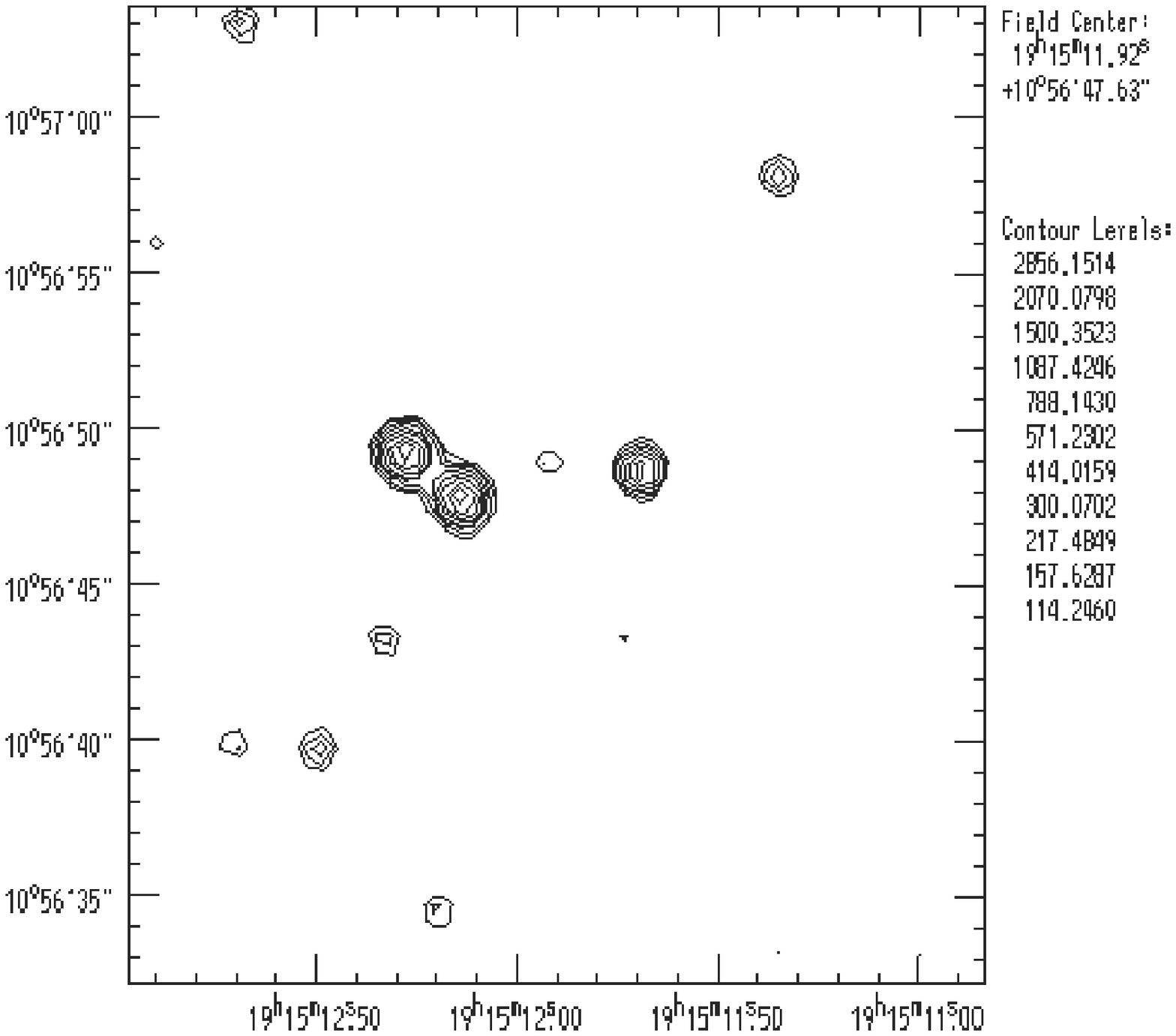}
\end{picture}

\caption{\label{grsimage} The figure on the left side is the original image of the field of view around $\grs$; the coordinates of $\grs$ are $\alpha(2000) = 19^{\rm h} 15^{\rm m} 11.^{\rm s}545$ and $\delta = 10^{\circ} 56^{'} 44.80 \arcsec$. The figure on the right side is the image of the field of view around $\grs$, with the PSF of $\grs$ being subtracted. We can not see any presence of jet structure, up to a magnitude in the K-band ($\lambda = 2.2 \, \mu$m) of m$_{K} > 17.5$ magnitude.}
\end{figure}

The image taken at la Silla is the median of 5 images exposed during 2 minutes. After taking each image of the object, an image of the sky was taken, to allow subtraction of the blank sky. The images are further treated by removal of the bias, the dark current, and the flat field. This image is shown in Fig. \ref{grsimage}. A point-spread-function (PSF) was synthesized thanks to the brightest stars of the field, as many as possible, and the less contaminated by others as possible, e.g. corresponding to the characteristics of good PSF stars (Massey \& Davis~\cite{mas92}). Scaling this typical PSF according to the stars surrounding the good PSF stars, we subtracted the PSF of these neighbouring stars: it remains only residuals of neighbouring stars around PSF stars. Therefore, we can reconstruct an improved PSF, synthesizing it thanks to these PSF stars, with only neighbours residuals. We repeat this mechanism until we obtain a good typical PSF. Finally, we scale this PSF, according to the PSF of $\grs$, and we subtract it to the infrared counterpart of $\grs$. The result is that we can not see any presence of jet structure, as the Fig. \ref{grsimage} shows it. The limit in magnitude that we can derive from our observations is m$_{K} > 17.5$ magnitude ($\lambda_{K} = 2.2 \, \mu$m).

We also searched for possible extended emission around the source $\grs$ by a multi-frequency analysis, thanks to a wavelet approach, and the result is identical.

To facilitate the search of a likely existing feature around $\grs$, we give in the Table \ref{movements} the various possible positions and configurations of the feature seen by Sams {\it et al.}~\cite{sam96}. The given ejection velocities are the minimum and maximum ejection velocities seen during the radio outbursts of $\grs$ (Mirabel {\it et al.}~\cite{mir96b}, Mirabel \& Rodríguez~\cite{mir94}), and the given expansion velocities are those usually seen towards the radio-emitting jets.

\begin{table}[t]
\caption{Various positions and configurations of the feature seen by Sams {\it et al.} (1996), according to its ejection and expansion velocity. \label{movements}}
\vspace{0.4cm}
\begin{center}
\begin{tabular}{|l|c|c|c|c|c|}
\hline
{\bf Ref.} & {\bf Date} & \multicolumn{2}{|c|}{\bf Center--Feature's distance} & \multicolumn{2}{|c|}{\bf Feature's FWHM} \\ \cline{3-6}
 & & \multicolumn{2}{|c|}{ $v_{ejection}$} & \multicolumn{2}{|c|}{ $v_{expansion}$} \\
 & & $1.0 \, c$ & $0.5 \, c$ & $0.5 \, c$ & $0.25 \, c$ \\ \cline{1-4}
\multicolumn{2}{|c|}{Ejection} & 26/Jun/1995 & 04/Jun/1995 & & \\ \hline
Sams~\cite{sam96} & 18-21/Jul/1995 & $0.3''$ & $0.3''$ & $0.25''$ & $0.25''$ \\
 & & $5.6 \times 10^{14} \metre$ & $5.6 \times 10^{14} \metre$ & $4.7 \times 10^{14} \metre$ & $4.7 \times 10^{14} \metre$ \\ \hline
Mirabel~\cite{mir96b} & 04/Aug/1995 & $0.53''$ & $0.42''$ & $0.48''$ & $0.37''$ \\
 & & $9.9 \times 10^{14} \metre$ & $7.8 \times 10^{14} \metre$ & $9.0 \times 10^{14} \metre$ & $6.8 \times 10^{14} \metre$ \\ \hline
Eikenberry~\cite{eik96} & 16-17/Oct/1995 & $1.55''$ & $0.92''$ & $1.5''$ & $0.87''$ \\
 & & $2.9 \times 10^{15} \metre$ & $1.7 \times 10^{15} \metre$ & $2.8 \times 10^{15} \metre$ & $1.6 \times 10^{15} \metre$ \\ \hline
\end{tabular}
\end{center}
\end{table}

Therefore, the infrared flux decreased by a factor $\geq 28$, in a time $\Delta t = 16.7 \mbox{ days}$: {\bf the lifetime of the near-infrared--emitting electrons} involved in this emission is: \[\tau < 17 \mbox{ days}\]

Sams {\it et al.}~\cite{sam96} point out the fact that the entire source has $K = 12.5 \mag$, saying that this makes it $0.5 \mag$ brighter than at any previously measured time. But Chaty {\it et al.}~\cite{cha96} have shown that $\grs$ exhibits some short- and long-timescale variability, and that the K-magnitude of $\grs$ was $12.15 \mag$ on 5 July 1994.

\section{The characteristics of the jet}

In this part, we see what are the implications of the detection and non-detection of the jet, if we suppose that the radiation seen in the near-infrared wavelengths is synchrotron emission. This gives constraints on the magnetic field in the ejected clouds, and on the Lorentz factor of the electrons ejected from the central source.

The total energy loss rate by synchrotron radiation is (Pacholczyk~\cite{pac72}): \[-(\frac{dE}{dt})_{\rm{synch}} (\ergs)= c_{2} B^{2} E^{2}\] where $c_{2} = 2.37 \times 10^{-3}$ cgs.

The near-infrared--emitting electrons lose all their energy in a time: \[\tau = -\frac{E}{(dE/dt)_{\rm{synch}}}\]

The critical frequency, for which these relativistic electrons emit most of the synchrotron radiation is (Pacholczyk~\cite{pac72}): \[\nu_{c} (\mbox{ Hz})= c_{1} B E^{2}\] where $c_{1} = 6.27 \times 10^{18}$ cgs.

From these equations, we can derive the magnetic field needed to produce the synchrotron radiation: \[B(\mbox{ G}) = 1.14 \, [\nu_{c}(10^{14} \mbox{ Hz})]^{-1/3} \, [\tau(\mbox{ days})]^{-2/3}\] and the energy of the electrons: \[E(\mbox{ erg}) = 3.73 \times 10^{-3} \, [\nu_{c}(10^{14} \mbox{ Hz})]^{2/3} \, [\tau(\mbox{ days})]^{1/3}\]

We know from the observations the radiative timescale of the electrons emitting the synchrotron radiation in the near-infrared wavelengths, especially in the K-band: $\tau < 17$ days; the critical frequency, which is here the frequency where we observe this synchrotron radiation, in the K-band at near-infrared wavelengths: $\nu_{c} = \nu(\mbox{ K}) = 1.36 \times 10^{14}$ Hz. We can therefore derive observational limits for the {\bf magnetic field in the ejected clouds:} \[B \gtsima 160 \mbox{ mG}\] and for the {\bf Lorentz factor of the internal motions} of the ultrarelativistic electrons in the ejected clouds: \[\gamma \ltsima 1.4 \times 10^{4}\] corresponding to an energy of $E = \gamma m c^{2} \ltsima 7.4 \GeV$.

These results are consistent with the previous measurements carried out during the 1994 radio outburst, with the polarization and assuming equipartition (Mirabel \& Rodríguez~\cite{mir96a}), which had led to a magnetic field $B \gtsima 50$ mG and a Lorentz factor $\gamma \sim 10^{3}$. Therefore, this suggests a magneto-hydrodynamic mechanism for the acceleration and collimation of the ejecta.

\section*{Acknowledgments}
We thank J.-L. Starck for the wavelet analysis of the images of $\grs$, using its software MOSAD, J. Martí and C. Gouiffes for helpful discussions.


\section*{References}
\begin{thebibliography}{99}

\bibitem{cha96}S. Chaty, I.F. Mirabel, P.-A. Duc, J.E. Wink, L.F. Rodríguez, \Journal{A\&A}{310}{825}{1996}.

\bibitem{eik96}S.S. Eikenberry and G.G. Fazio, \Journal{ApJL}{475}{L53}{1997}.

\bibitem{mas92}P. Massey and L.E. Davis in {\em A User's Guide to Stellar CCD Photometry with IRAF} (1992).

\bibitem{mir94}I.F. Mirabel and L.F. Rodríguez, \Journal{Nat}{371}{46}{1994}.

\bibitem{mir96a}I.F. Mirabel and L.F. Rodríguez, \Journal{BAAS}{189}{5801}{1996}.

\bibitem{mir96b}I.F. Mirabel, L.F. Rodríguez, S. Chaty {\it et al.}, \Journal{ApJ}{472}{L111}{1996}.

\bibitem{pac72}A.G. Pacholczyk in {\em Radio Astrophysics}, ed. W.H. Freeman and Co. (San Francisco, 1972).

\bibitem{sam96}B.J. Sams, A. Eckart, R. Sunyaev, \Journal{Nat}{382}{47}{1996}

\end{thebibliography}
\end{document}